\documentclass[12pt,oneside,onecolumn,floatfix]{article}
\usepackage{epsfig}
\usepackage{amsmath,amssymb,amsfonts}
\usepackage{a4wide}
\usepackage[top=30pt,bottom=30pt,left=48pt,right=46pt]{geometry}
\usepackage{slashed, enumitem}
\usepackage[utf8]{inputenc}
\usepackage{jcappub}
\usepackage{enumerate}
\usepackage{float,color}
\usepackage{subfigure}
\usepackage{multirow}
\usepackage{slashed}
\usepackage{placeins}
\usepackage{wasysym} 
\usepackage{mathrsfs} 
\usepackage{caption}
\usepackage{amsmath}
\usepackage{amssymb}
\usepackage{graphicx}
\usepackage{slashed}
\usepackage{multirow}
\usepackage{placeins}
\usepackage[dvipsnames]{xcolor}
\usepackage{epstopdf}
\usepackage{soul}
\usepackage{tikz}
\usepackage[capitalise, english]{cleveref}
\usepackage{siunitx}
\usepackage{xspace}
\usepackage{booktabs}
\usetikzlibrary{trees}
\usetikzlibrary{decorations.pathmorphing}
\usetikzlibrary{decorations.markings}
\usepackage[colorlinks=true,citecolor=blue,linkcolor=blue]{hyperref}

\definecolor{americanrose}{rgb}{1.0, 0.01, 0.24}
\definecolor{ao}{rgb}{0.0, 0.0, 0.97}
\definecolor{myviolet}{rgb}{0.83, 0.11, 0.83}

\newcommand\myshade{80}
\colorlet{mylinkcolor}{ao}
\colorlet{mycitecolor}{americanrose}
\colorlet{myurlcolor}{ao}

\hypersetup{
	linkcolor  = mylinkcolor!\myshade!black,
	citecolor  = mycitecolor!\myshade!black,
	urlcolor   = myurlcolor!\myshade!black,
	colorlinks = true
}

\def \beq{\begin{equation}}
	\def \eeq{\end{equation}}
\def \bea{\begin{eqnarray}}
	\def \eea{\end{eqnarray}}
\def \ba{\begin{array}}
	\def \ea{\end{array}}

\usepackage{tikz,xcolor,hyperref}

\definecolor{notecolor}{rgb}{0.8,0,0}
\definecolor{green}{rgb}{0.0, 0.5, 0.0}

\definecolor{lime}{HTML}{A6CE39}
\DeclareRobustCommand{\orcidicon}{\hspace{-3mm}
	\begin{tikzpicture}
		\draw[lime, fill=lime] (0,0) 
		circle [radius=0.16] 
		node[white] {\hspace{0.1mm}{\fontfamily{qag}\selectfont \tiny ID}};
		\draw[white, fill=white] (-0.07,0.1) 
		circle [radius=0.01];
	\end{tikzpicture}
	\hspace{-5mm}
}

\foreach \x in {A, ..., Z}{\expandafter\xdef\csname orcid\x\endcsname{\noexpand\href{https://orcid.org/\csname orcidauthor\x\endcsname}
		{\noexpand\orcidicon}}
}

\begin{document}

\title{Neutrinos from Earth-Bound Dark Matter Annihilation}

\author[a,b]{Maxim Pospelov}
\emailAdd{pospelov@umn.edu}
\affiliation[a]{School of Physics and Astronomy, University of Minnesota, Minneapolis, MN 55455, USA}
\affiliation[b]{William I. Fine Theoretical Physics Institute, School of Physics and Astronomy, University of Minnesota, Minneapolis, MN 55455, USA}

\author[a,c]{and Anupam Ray\orcidB{}}
\emailAdd{anupam.ray@berkeley.edu}
\affiliation[c]{Department of Physics, University of California Berkeley, Berkeley, California 94720, USA}

\date{\today}


\begin{abstract}
 {A sub-component of dark matter with a short collision length compared to a planetary size leads to efficient accumulation of dark matter in astrophysical bodies. We analyze possible neutrino signals from the annihilation of such dark matter and conclude that in the optically thick regime for dark matter capture, the Earth provides the largest neutrino flux. Using the results of the existing searches, we consider two scenarios for the neutrino flux, from stopped mesons and prompt higher-energy neutrinos. In both cases we exclude some previously unexplored parts of the parameter space (dark matter mass, its abundance, and the scattering cross section on nuclei) by recasting the existing neutrino searches. } 
\end{abstract}
\subheader{N3AS-23-030}

\maketitle
\section{Introduction}
Unambiguous evidence of a non-baryonic form of matter that seeds cosmic structures, commonly known as dark matter (DM), constituting $\sim$ 27\% of the total energy density
of the Universe. Its existence finds plenty of evidence through cosmological and astrophysical observations~\cite{Aghanim:2018eyx}. Despite extensive searches over the last few decades of its non-gravitational manifestations, DM remains mysterious. In the absence of an irrefutable signal, these terrestrial and astrophysical searches are placing stringent exclusions on non-gravitational interactions of DM with the ordinary baryonic matter over a wide mass range~\cite{Cooley:2022ufh,Baryakhtar:2022hbu,Boddy:2022knd}.

DM accumulation in stellar objects is yet another promising astrophysical probe of DM interactions~\cite{1985ApJ...296..679P,Gould:1987ir,Gould:1987ju}. Because of the non-gravitational interaction of dark matter with the baryonic matter, DM particles from the galactic halo can down-scatter to energies below the local escape energy, and become gravitationally bound to the stellar objects. These bound DM particles lose more energy via repetitive scatterings with the stellar material, and eventually thermalize inside the stellar volume. Such bound thermalized DM particles can be copiously present inside the stellar volume, and they have interesting phenomenological signatures (see \textit{e.g.,} Refs.~\cite{Gould:1989hm,Mack:2007xj,Ray:2023auh,Bhattacharya:2023stq,Nguyen:2022zwb,Bramante:2023djs}). 

Of particular importance is the signal of the DM annihilation to the Standard Model (SM) states in the Sun and the Earth, that may manifest itself in a variety of different ways. In particular, annihilation may result in the flux of energetic neutrinos associated with the direction to the center of the Sun/Earth~\cite{Silk:1985ax,Srednicki:1986vj}. In some models, the existence of meta-stable non-SM intermediate mediator states may take the products of annihilation to astronomical distances before mediators decay producing visible SM particles \cite{Batell:2009zp,Schuster:2009au}. Finally, it has been recently pointed out that in some models with efficient trapping of the $\mathcal{O}({\rm GeV})$ scale DM particles, the annihilation directly into the SM states inside the active volumes of the neutrino detector also represent a viable option \cite{McKeen:2023ztq}. 

In this work, we primarily consider DM particles that are efficiently trapped by the Earth, which occurs when the collision length is small compared to the Earth's size, $\ell_{\rm col} \ll R_\oplus$. This in turn results from  the DM cross section on ordinary atoms being much larger than benchmark values due to {\em e.g.} electroweak force. Due to the existing constraints, large cross section implies that such DM particles have to be a {\em sub-component} of DM, {\em i.e.} not fully saturating the DM abundance. 
We define $\chi$ as a DM sub-species that makes up a fraction $f_{\chi}$ $(f_{\chi}=\rho_{\chi}/\rho_{\rm{DM}} \leq 1)$ of the present day dark matter density, and has a sizeable scattering cross-section $\sigma_{\chi n}$ with the nucleons. Efficient trapping causes $\chi$ particles to be enormously abundant inside the Earth volume, even if their cosmological abundance is tiny ($f_{\chi} \ll1$). In fact, it is well appreciated that owing to the enormous size of the Earth and cosmologically long lifetime, their terrestrial density may be enhanced $\mathcal{O} (15)$ orders of magnitude over the local Galactic DM density~\cite{Neufeld:2018slx,Pospelov:2020ktu,Berlin:2023zpn}. Furthermore, if the $\chi$ particles are sufficiently light, they  distribute over the  entire Earth volume, rather than concentrating towards the core, making their surface-density tantalizingly large, up to $f_{\chi} \times 10^{14}$ cm$^{-3}$ for DM mass of 1 GeV.
However, since these $\chi$ particles carry a minuscule amount of
kinetic energy $\sim kT =0.03$ eV, they are almost impossible to detect in the traditional direct detection experiments as these experiments primarily rely on elastic scattering signatures. A few recent studies have proposed their detection via up-scattering them through collisions~\cite{McKeen:2022poo}, by utilizing low threshold quantum
sensors~\cite{Budker:2021quh,Das:2022srn,Billard:2022cqd}, and more importantly, via searching their annihilation signatures inside large-volume neutrino detectors, such as, Super-Kamiokande(SK)~\cite{McKeen:2023ztq}. 

\begin{figure}
	\centering
	\includegraphics[width=0.97\textwidth]{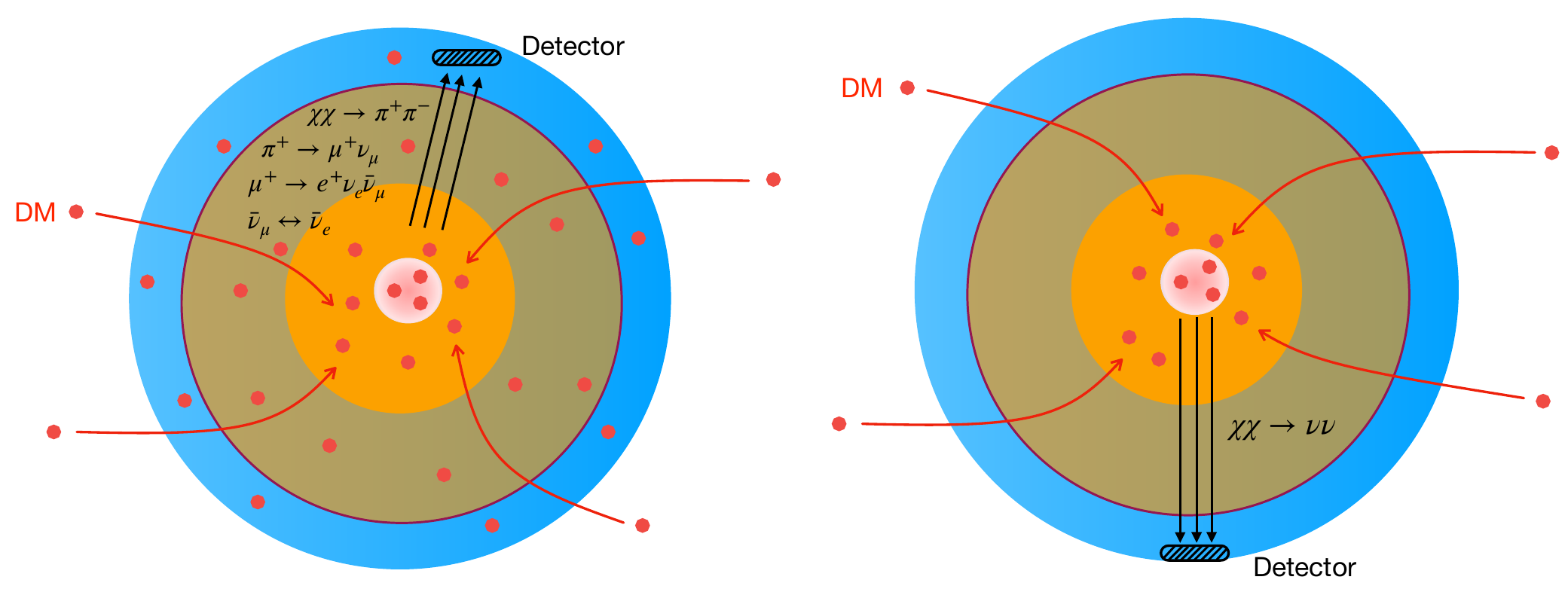}
	\caption{Annihilation of Earth-bound DM into neutrinos is schematically illustrated. In the left panel, low energy neutrinos are produced via stopped pion decay, whereas, in the right panel, high energy neutrinos  are produced via direct annihilation of Earth-bound DM particles.}
\label{schematic}
\end{figure}

Here, we examine the detection of $\chi$ particles via their annihilation into neutrinos. Despite the fact that the rate of neutrino interaction is rather small at sub-electroweak scale energies,  the neutrinos have an obvious advantage of taking the signal to a long distance.  We consider two phenomenological scenarios: $\chi$ annihilation to light mesons (pions) that stop and decay at rest that limits neutrino energies to 50 MeV,  and direct annihilation of $\chi$ to neutrinos (or to other intermediate particles that decay directly to neutrinos before stopping), 
that creates neutrino flux with energies of $O(m_\chi)$. We find that in the first case, 
that current generation of neutrino experiments (mostly Super-Kamiokande) can constrain $f_\chi$ down to  $\sim 10^{-4}$ with sub-electoweak masses for $m_\chi$. The second case, direct annihilation to energetic neutrinos, followed by their detection at IceCube DeepCore can provide sensitivity to very small $f_\chi$ ($f_\chi \sim 10^{-8}$). We provide a schematic diagram of these two scenarios in Fig.\,\ref{schematic}.

It is important to note that in Ref.~\cite{McKeen:2023ztq}, \textit{local} annihilation of $\chi$ particles inside the fiducial volume of large neutrino detectors, such as Super-Kamiokande, has been utilized to provide world-leading exclusions on DM interactions. However, this method is not well-suited for probing interactions of  relatively heavy DM particles with $m_{\chi} \geq$\,5\,GeV. This is simply because, with increase in DM mass, $\chi$ particles shrink more towards the Earth-core, resulting in a negligible number of dark matter particles inside the fiducial volume of Super-Kamiokande. Therefore, in order to probe the interactions of heavier dark matter particles, analyzing the neutrino signals from Earth-bound DM annihilation appears promising.

The rest of the paper is organized as follows. In Sec.\,\ref{flux_neutrinos}, we estimate the flux of neutrinos from Earth-bound DM annihilation, demonstrating that Earth as the most optimal detector for this purpose. In Sec.\,\ref{results}, we present our results for two phenomenological scenarios: low energy neutrinos from stopped pion decay as well as direct annihilation to energetic neutrinos and their detection at Super-Kamiokande \& IceCube DeepCore, respectively. We summarize and conclude in Sec.\,\ref{conclusion}.

\section{Flux of Neutrinos from Earth-bound DM Annihilation}\label{flux_neutrinos}
The total number of $\chi$ particles inside the Earth volume is given by 
\begin{align}
	\label{Nchi}
	\frac{dN_\chi}{dt} = \Gamma_{\rm cap} - N_{\chi}\Gamma_{\rm evap} - N_\chi^2\Gamma_{\rm ann}\,,
\end{align}
where $\Gamma_{\rm cap}$, $\Gamma_{\rm evap}$, and $\Gamma_{\rm ann}$ denotes the capture, evaporation, and annihilation rate. In the following, we briefly discuss each of these rates.

\begin{figure}
	\centering
	\includegraphics[width=0.65\textwidth]{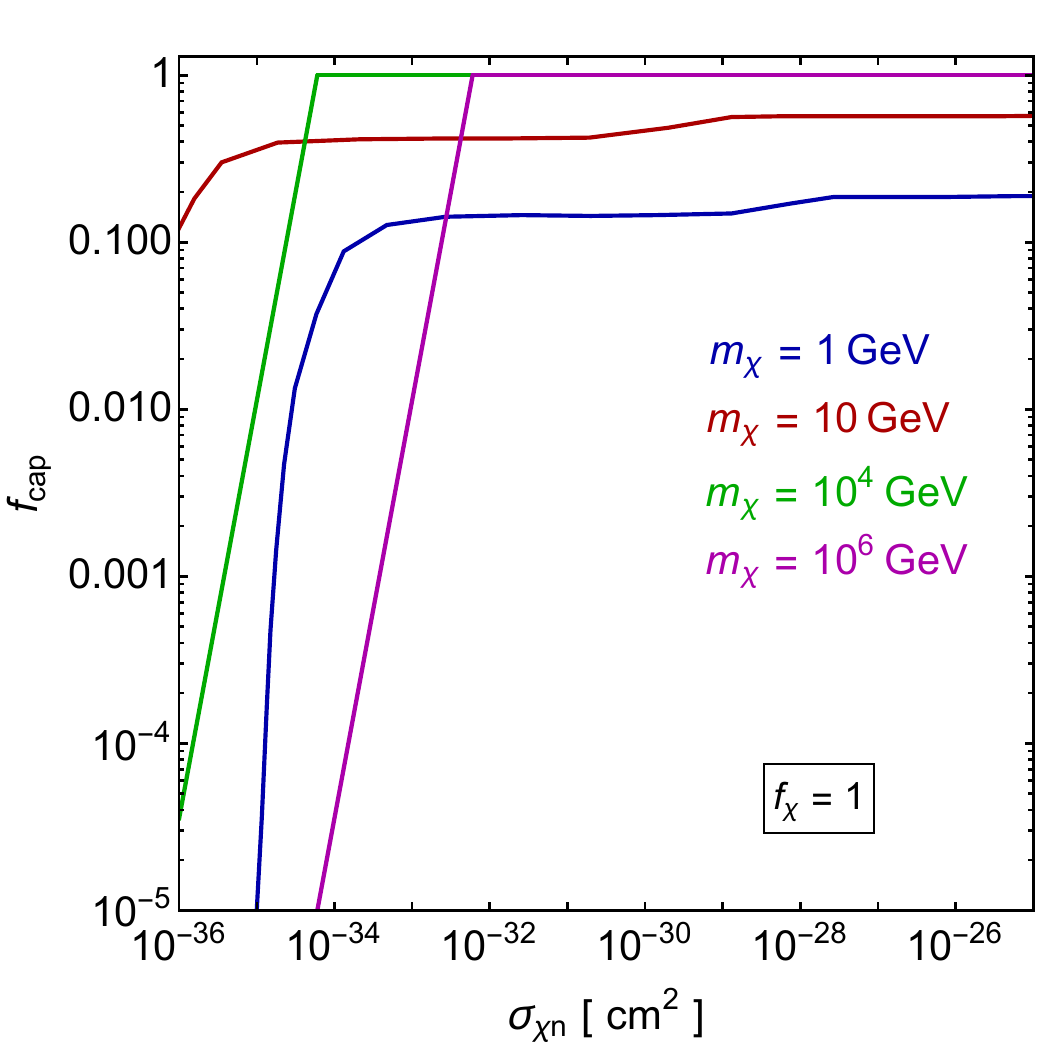}
	\caption{Capture fraction for the Earth is shown for our parameter range of interest. For DM mass of 1$-$10 GeV, we obtain $f_{\rm cap}$ by using the recent numerical simulation results~\cite{Bramante:2022pmn}, whereas, for heavier DM ($m_{\chi} \gg 10$ GeV), we use the results from Ref.~\cite{Acevedo:2019agu}.}
	\label{fcapp}
\end{figure}

Beginning with $\Gamma_{\rm cap}$, we define the maximal capture rate as geometric capture rate $(\Gamma_{\rm geo})$, which occurs when all of the $\chi$ particles that transit through the Earth get captured. Depending on the DM-nucleon scattering cross-section and DM mass, only a certain fraction $(f_{\rm cap})$ of the geometric capture rate $(\Gamma_{\rm geo})$ gets trapped, and we quote it as capture fraction. In the following, we briefly discuss about capture fraction for the Earth (for clarity, in Fig.\,\ref{fcapp}, we show $f_{\rm cap}$ for our parameter range of interest). For large DM-nucleon scattering cross-section, which is of primary interest here, $f_{\rm cap}$ behaves differently for heavier and lighter DM.  For lighter DM ($m_\chi \leq m_A$, where $m_A$ is a typical atomic mass), the probability of reflection is quite significant and hence, the capture fraction never reaches unity, i.e., $\Gamma_{\rm cap} < \Gamma_{\rm geo}$~\cite{Bramante:2022pmn}. Whereas, for heavier DM, capture fraction can reach unity if the DM-nucleon scattering cross-section is significantly large, i.e.,  $\Gamma_{\rm cap} \approx \Gamma_{\rm geo}$. Of course, this occurs when the available number of scatterings, which is dictated by the DM-nucleon scattering cross-section, is larger than the required number of scatterings to stop the DM particles. On the other hand, for small DM-nucleon scattering cross-section (optically thin regime), capture fraction is sufficiently small ($f_{\rm cap} \ll 1$), and the capture rate is always much less than the geometric capture rate. We use the recent numerical simulations~\cite{Bramante:2022pmn} to estimate the value of  $f_{\rm cap}$, which agrees reasonably well with the previous analytical estimate in Ref.~\cite{Neufeld:2018slx}.  Quantitatively, it suggests that for DM-nucleon scattering cross-section of $[10^{-34},10^{-26}]$ cm$^2$, and for DM mass of 1 GeV, $f_{\rm cap} \sim 0.1$, reducing further as $\sqrt{m_{\chi}}$ with lower DM masses. For heavier DM capture, i.e., $m_{\chi} \gg 10$ GeV, we use the capture fraction from Ref.~\cite{Acevedo:2019agu}.
To summarize, we use the following capture rate for the optically thick regime
\begin{align}
	\Gamma_{\rm cap}  =f_{\rm cap} \times \Gamma_{\rm{geom}} =f_{\rm cap} \times \sqrt{\frac{8}{3 \pi}} \frac{f_{\chi} \rho_{\rm DM}v_{\rm gal}}{m_{\chi}} 
	\times \pi R^2_{\oplus} \,,
\end{align}
where $R_{\oplus}$ is the radius of the Earth, $\rho_{\rm DM} = 0.4\,\rm{GeV}\,\rm{cm}^{-3}$ denotes the local Galactic DM density, and $v_{\rm gal} = 220$ km/s denotes the typical velocity of the $\chi$ particles in the Galactic halo.

In order to determine the evaporation and annihilation rates, we need to estimate the spatial distribution of the $\chi$ particles inside the Earth-volume.  The number density of bound $\chi$ particles inside the Earth-volume, $n_{\chi} (r)$, is essentially governed by the Boltzmann equation that combines the effects of gravity, concentration diffusion, and thermal diffusion~\cite{Gould:1989hm,Leane:2022hkk}
\begin{align}
		\label{diff}
	\frac{\nabla n_{\chi}(r) }{n_{\chi}(r)}+ \left(\kappa +1 \right) 	\frac{\nabla T(r) }{T(r)}+\frac{m_{\chi} g(r)}{k_BT(r)} = 0\,.
\end{align}
In the above equation, we have used the
hydrostatic equilibrium criterion as the diffusion timescales for the $\chi$ particles are short as compared to the other relevant timescales.  $\kappa \sim -1/\left[2(1+m_{\chi}/m_{A})^{3/2}\right]$ denotes the thermal
diffusion co-efficient~\cite{Leane:2022hkk} and $T(r)$ denotes the temperature profile of the Earth. For the temperature and density profiles
of the Earth, we follow Refs.~\cite{Dziewonski:1981xy,https://doi.org/10.1002/2017JB014723}. We also define a dimensionless radial profile function, $G_\chi(r)=V_\oplus n_\chi (r) \big/ \int^{R_{\oplus}}_{r=0} 4 \pi r^2 n_{\chi} (r) dr$, such that, for uniform distribution of $\chi$ particles, the profile function is trivial, \textit{i.e.},  $G_\chi(r) =1$. Distribution function $G_{\chi}(r)$ does not depend on the total number of trapped particles $N_\chi$ and therefore can be evaluated separately from Eq. (\ref{Nchi}).

By solving  Eq.~(\ref{diff}) for $n_{\chi} (r)$,
we find that for light DM masses  the density
profile is relatively constant and mildly increases toward the Earth-core. For heavier $m_{\chi}$,  $\chi$ particles
shrink toward the Earth-core, leading to a substantial depletion of the surface density, and approaches a Dirac-delta distribution at the center of the Earth for $m_{\chi} \to \infty$.

As is well known, thermal fluctuations can bring light DM particles above the escape velocity from the Earth
and open the loss channel commonly referred to as thermal evaporation of dark matter particles. If the collision length of dark matter particles is small, the evaporation will effectively occur from the ``last scattering surface", and we estimate the evaporation of $\chi$ particles by adopting the Jeans’ expression~\cite{Neufeld:2018slx}
\begin{align}
	\Gamma_{\rm evap} = G_\chi(R_{\rm LSS})\times \frac{3R_{\rm LSS}^2}{R_\oplus^3}\times \frac{v_{\rm LSS}^2+v_{\rm esc}^2}{2\pi^{1/2}v_{\rm LSS}} \exp(- v_{\rm esc}^2/v_{\rm LSS}^2),
\end{align}
where $R_{\rm LSS}$ and $v_{\rm LSS}$ denotes the radius and DM thermal velocity at the last scattering surface, respectively.We obtain the ``last scattering surface" via
\begin{equation}
	\int_{R_{\rm LSS}}^{\infty} dr \sum_j \sigma_{\chi j} n_j(r) =1\,,
\end{equation}
where $\sigma_{\chi j}$ denotes the scattering cross-section between DM and the $j$-th nuclei and $n_j(r)$ denotes the number density of the $j$-th nuclei. We use Preliminary Reference Earth Model for the density profile of the Earth~\cite{Dziewonski:1981xy}, and for the chemical compositional profile, we use Table I of Ref.~\cite{Bramante:2019fhi}. For the density, temperature, and compositional profile of the Earth's atmosphere, we use NRLMSISE-00 model~\cite{2002JGRA..107.1468P}. We note that, for large scattering cross sections, of primary interest here, last scattering surface lies near the Earth-surface or in the atmosphere, i.e., $R_{\rm LSS} \simeq R_\oplus$. 
Quantitatively, the effect of evaporation is always negligible for $\chi$ particles heavier than 10 GeV, and is  significant for $m_\chi \gtrsim 1$\,GeV irrespective of the DM-nucleon scattering cross-section~\cite{1990ApJ...356..302G,Garani:2017jcj,Bramante:2022pmn,Garani:2021feo}.

Finally, the annihilation rate of $\chi$ particles (for energy-independent $s$-wave annihilation) is given by
\begin{equation}
	\Gamma_{\rm ann} = \frac{4\pi}{N_\chi^2} \int_0^{R_\oplus} r^2dr n^2_\chi(r)\langle \sigma v \rangle_{\rm ann}	= \frac{4\pi\langle \sigma v \rangle_{\rm ann}}{V_\oplus^2} \int_0^{R_\oplus} r^2dr G^2_\chi(r)\,.
\end{equation}
Combining each of these rates, we can integrate Eq.~(\ref{Nchi}) to solve for $N_{\rm \chi}$. We note that for most of the parameter space relevant to our problem, dynamical equilibrium is achieved $(dN_\chi/dt = 0)$, implying either the annihilation or the evaporation rate counter-balances the capture rate. In this scenario, it is straightforward to solve Eq.~(\ref{Nchi}) to obtain~\cite{McKeen:2023ztq} 
\begin{equation}
2N_\chi = \left[(\tau_{\rm ann}/\tau_{\rm evap})^2+4\Gamma_{\rm cap}\tau_{\rm ann} \right]^{1/2}-\tau_{\rm ann}/\tau_{\rm evap} \,.
\end{equation}

\begin{figure}
	\centering
	\includegraphics[width=0.65\textwidth]{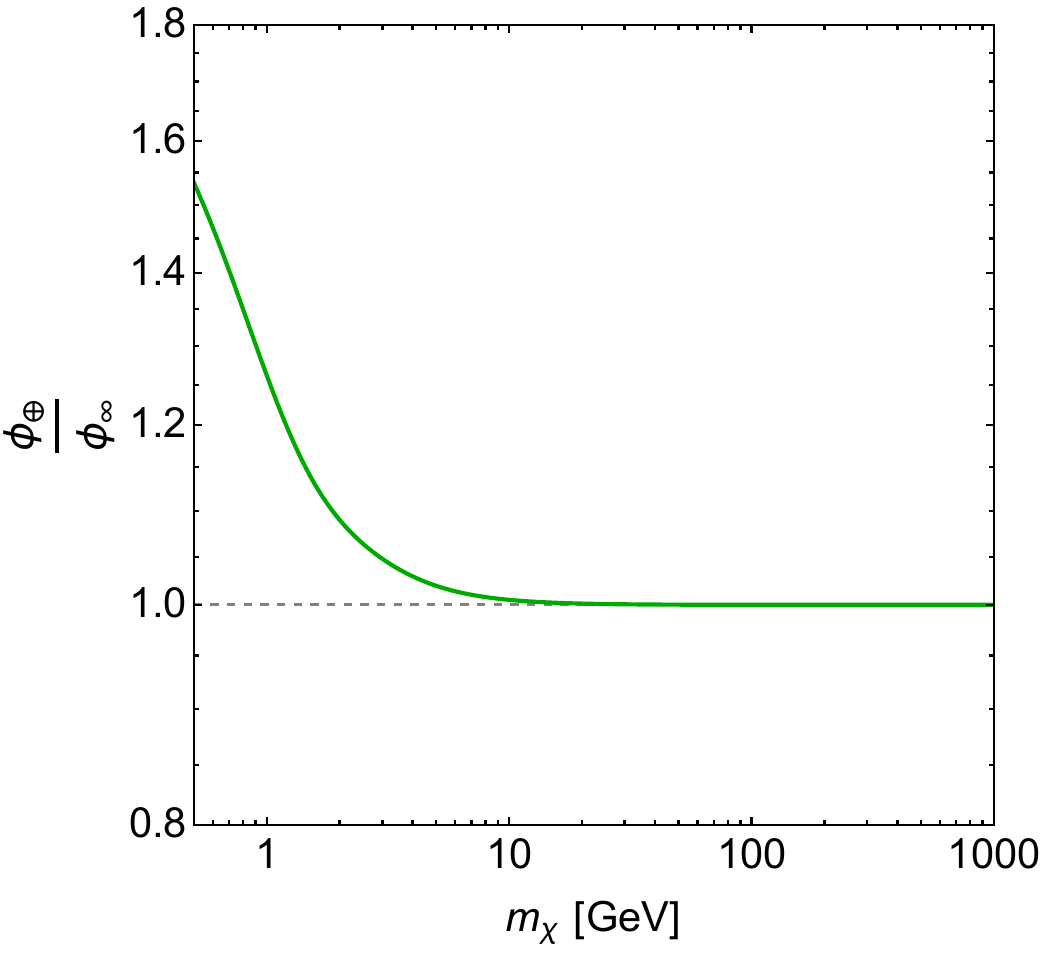}
	\caption{Flux of neutrinos from annihilation of Earth-bound DM particles $(\phi_{\oplus})$ as compared to the scenario when all the Earth-bound particles reside exactly at the center of the Earth ($\phi_{\infty}$). With increase in mass, $(\phi_{\oplus})$ gradually approaches to $\phi_{\infty}$.}
	\label{los}
\end{figure}

When the DM is heavy, all annihilations occur close to the center, and the flux of neutrinos at Earth's surface, resulting from the annihilation events can be calculated by dividing the total annihilation rate by $4\pi R_\oplus^2$. Lighter dark matter gives spatial extent to the region where annihilations occur, that can be comparable to Earth's size. 
In other words, spatial distribution of annihilation creates a correction to the total flux of annihilation products ({\em i.e.} neutrinos) reaching a detector. 
If evaporation can be neglected, one can define the total flux of the neutrinos from annihilation of $\chi$ particles. Assuming that single annihilation event produces $N_{\nu}$ neutrinos, we obtain the expression for the neutrino flux as 
\begin{equation}
	\label{flux}
	\phi_{\oplus} =N_\nu \times  \int \frac{\Gamma_{\rm cap} n^2_{\chi}(r)r^2dr d\Omega}{4 \pi \left(R^2_{\oplus}+r^2-2R_{\oplus}r\cos\theta\right)}\frac{1}{\int^{R_{\oplus}}_{r=0}4 \pi r^2n^2_{\chi}(r) dr}\,,
\end{equation}
where $d\Omega= 2\pi \sin \theta d\theta$ with the angular integral runs from $\theta = 0$ to $\theta = \pi$. Note that, in the limit of $m_{\chi} \to \infty$, $\chi$ particles concentrate at the center of the Earth, and we recover the familiar expression of $\phi_{\infty}/N_\nu  = \Gamma_{\rm cap}/ (4 \pi R^2_{\oplus})$ from Eq.~(\ref{flux}). In Fig.\,\ref{los}, we show that with increase in $m_{\chi}$, $\phi_{\oplus}$ gradually approaches $\phi_{\infty}$, whereas, for light DM masses $\phi_{\oplus} > \phi_{\infty}$. 
\medskip\\
\textbf{Earth as the most optimal detector:} In most of the WIMP dark matter models, the indirect dark matter detection via annihilation to neutrinos is dominated  by annihilations inside the Sun. This is typically the case for the optically thin regime, as larger number of target nuclei lead to a far greater accumulation rate of dark matter, scaling with the mass of an astronomical body. It is important to stress that, in the optically thick regime (large DM-nucleon scattering cross-section), Earth is the most optimal ``collector'' of DM interactions resulting to a comparatively larger neutrino annihilation signal. As compared to the Sun, Earth accumulates significantly fewer number of $\chi$ particles, but in this case the capture of particles is proportional to the surface area
of a planet/star. Together with the much larger Earth-Sun distance, this makes the flux of the Earth-bound DM considerably larger than that from the Solar-bound DM. Quantitatively, for a DM mass of 10 GeV, we find that flux of the Earth-bound DM is $\sim 4000$ times larger than the flux of Solar-bound DM particles. As compared to the Jupiter-bound DM, the flux enhancement is even larger; by a factor of $\sim 10^8$.
Therefore, our focus on the annihilation to neutrinos in the center of the Earth is well justified.

\section{Results}\label{results}
\begin{figure*}
	\centering
	\includegraphics[width=0.32\textwidth]{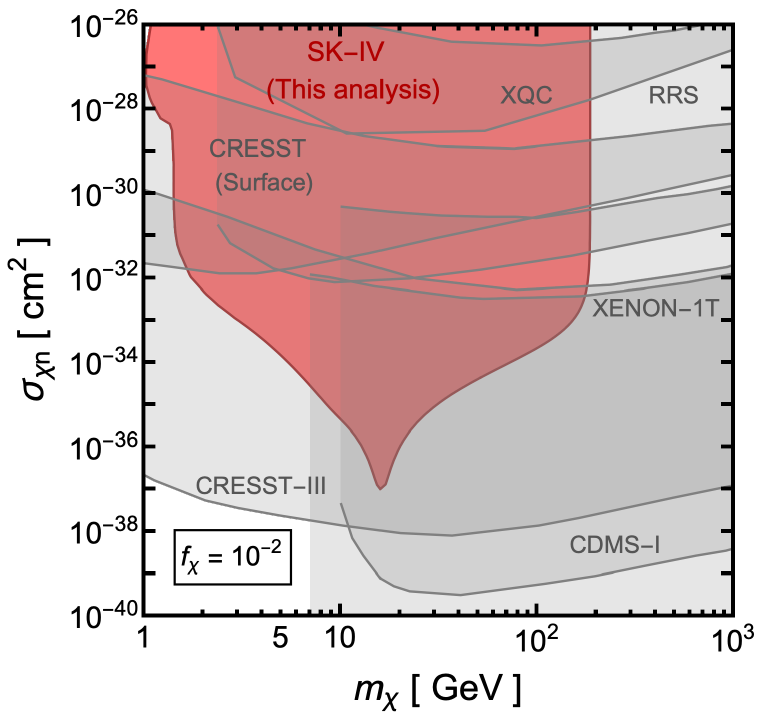}
		\hspace*{0.05 cm}
	\includegraphics[width=0.32\textwidth]{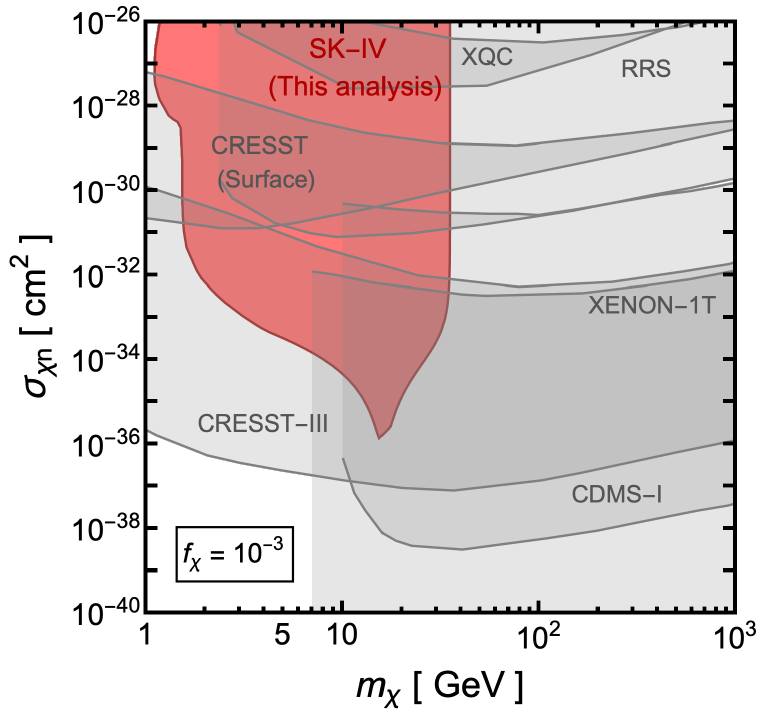}
		\hspace*{0.05 cm}
	 \includegraphics[width=0.32\textwidth]{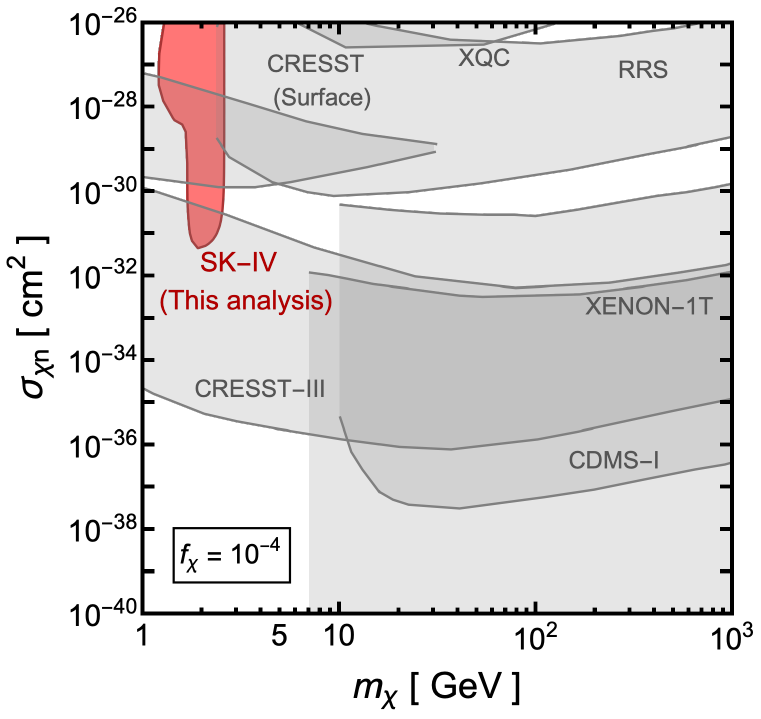}\vspace{0.5 cm}  
	 \includegraphics[width=0.32\textwidth]{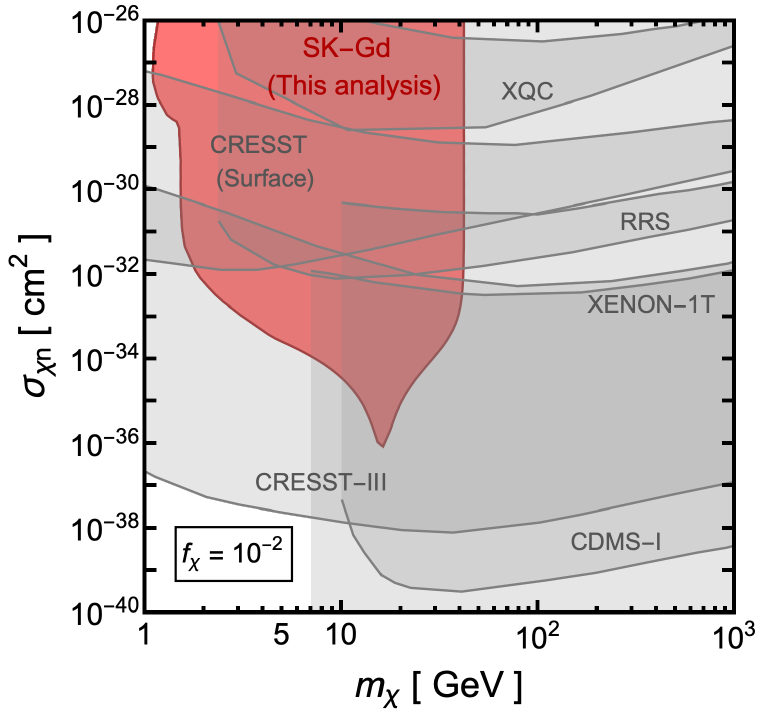}
	 \hspace*{0.05 cm}
	 \includegraphics[width=0.32\textwidth]{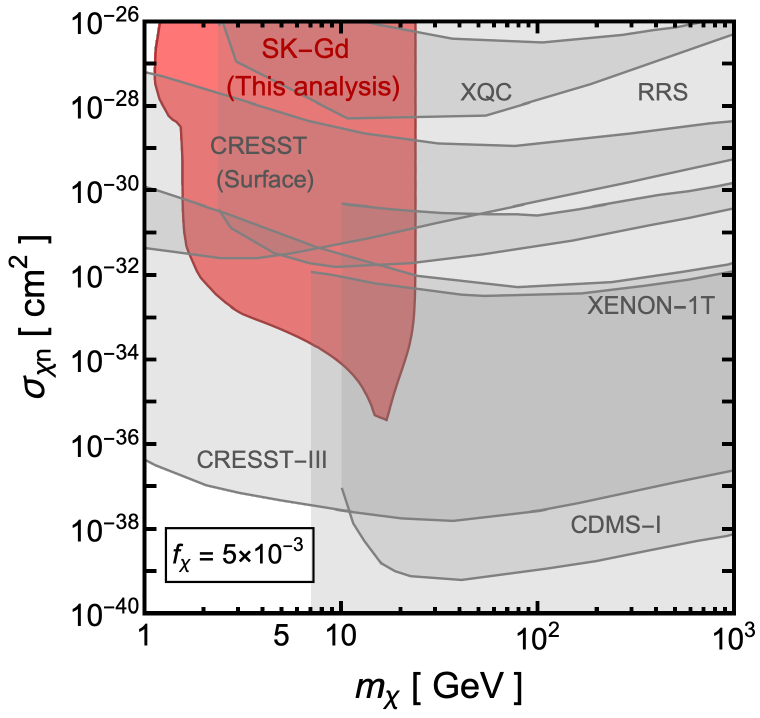}
	 \hspace*{0.05 cm}
	 \includegraphics[width=0.32\textwidth]{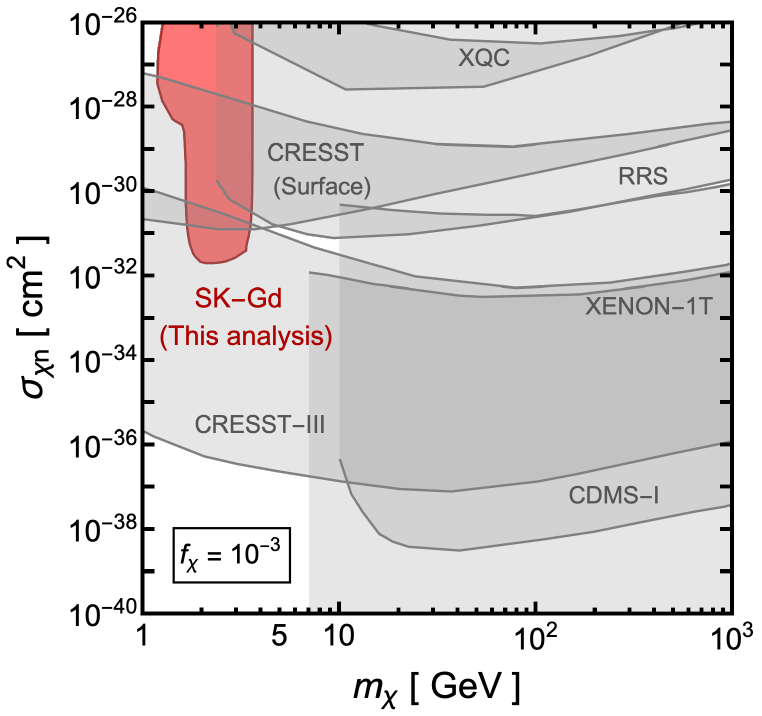}
	\caption{Constraints on DM-nucleon scattering cross-section (red shaded regions) from Earth-bound DM annihilation. We use the flux upper limits inferred from non-detection of diffuse supernovae background searches at Super-Kamiokande detector to derive the constraints. In the top panels, we use the SK result with pure-water (22.5 $\times$ 2970 kton-day)~\cite{Super-Kamiokande:2021jaq}, whereas, in the bottom panels, we use the SK result with  0.01wt\% gadolinium loaded water (22.5 $\times$ 552.2 kton-day)~\cite{Super-Kamiokande:2023xup}. The existing exclusion limits from underground as well as surface detectors (gray shaded regions) including CRESST-III~\cite{CRESST:2019jnq}, CRESST surface~\cite{CRESST:2017ues}, XENON-1T~\cite{XENON:2018voc}, CDMS-I~\cite{CDMS:2002moo}, and high-altitude detectors (RRS~\cite{Rich:1987st}, XQC~\cite{Erickcek:2007jv}) are also shown for comparison. }
	\label{lowenergy}
\end{figure*}

\begin{figure*}
	\centering
	\includegraphics[width=0.32\textwidth]{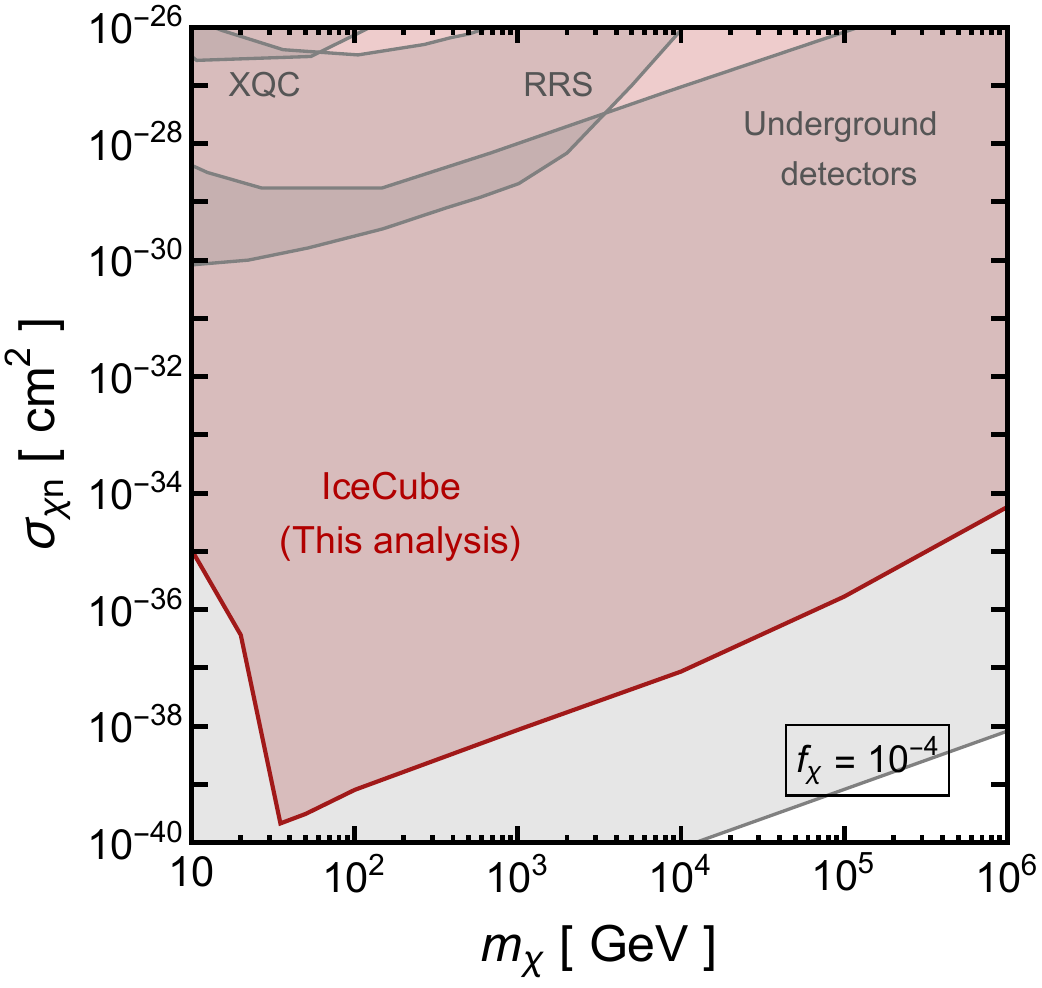}
	\hspace*{0.05 cm}
	\includegraphics[width=0.32\textwidth]{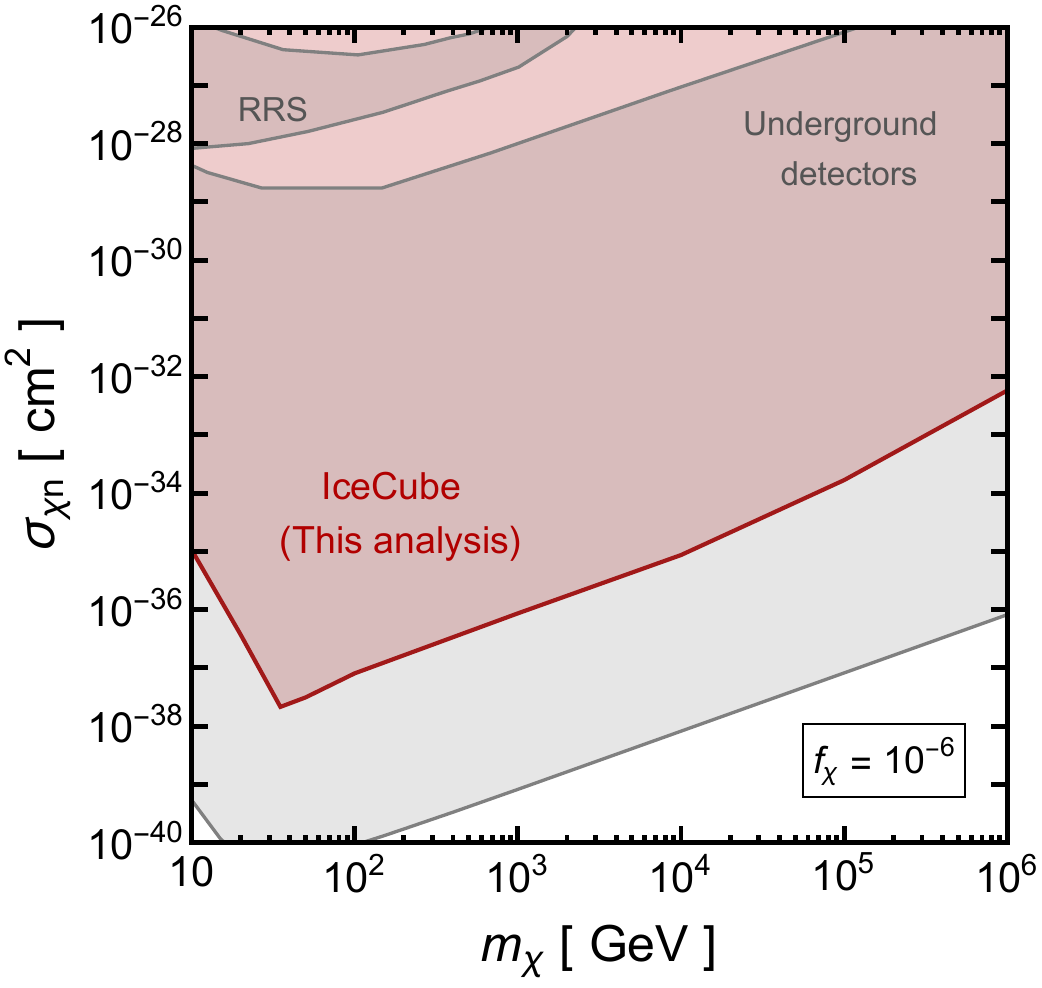}
	\hspace*{0.05 cm}
	\includegraphics[width=0.32\textwidth]{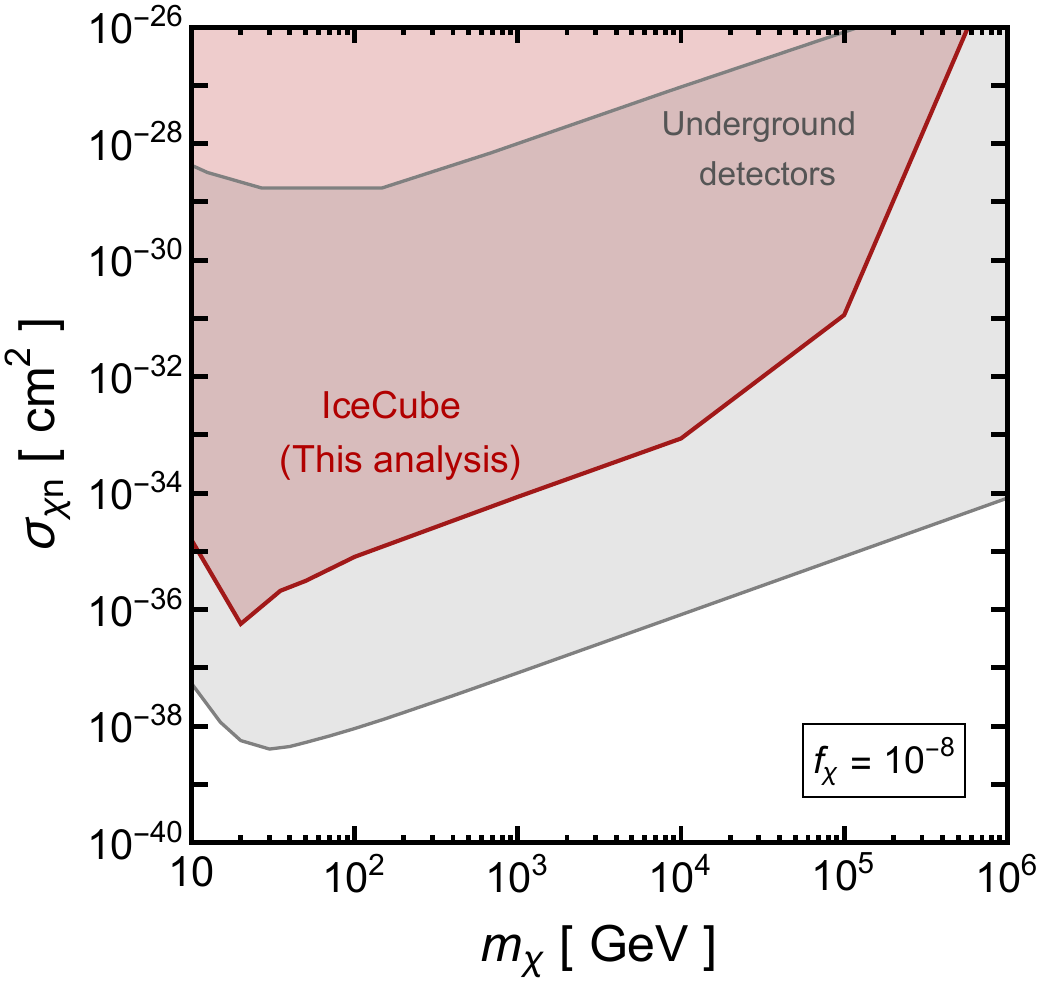}
	\caption{Constraints on DM-nucleon  scattering cross-section (red shaded regions) from direct annihilation of $\chi$ particles ($\chi \chi \to \nu \bar{\nu}$) from the center of the Earth. We use the IceCube DeepCore upper limits on DM annihilation rate for $\chi \chi \to \nu \bar{\nu}$, which is obtained by using a 6.75 years data~\cite{IceCube:2021xzo}. Existing constraints from underground as well as surface detectors (which includes XENON-1T, CDMS-I, CRESST-III, and CRESST-surface; collected in~\cite{Kavanagh:2017cru,Digman:2019wdm,Carney:2022gse}) and high-altitude detectors (RRS~\cite{Rich:1987st}, XQC~\cite{Erickcek:2007jv}) are also shown for comparison. }
	\label{highenergy}
\end{figure*}
\subsection{Low Energy Neutrinos from Stopped Meson Decay}
DM annihilation may result in neutrino fluxes of vastly different energies. 
In this subsection, we concentrate on sub-electroweak scale dark matter that annihilates into light meson states ({\em e.g.} pions) that mostly stop due to Coulomb and hadronic interactions and then decay at rest~\cite{Rott:2012qb,Bernal:2012qh}. Such scenario is especially relevant if the DM-nucleon interaction is 
mediated by a dark photon with $2m_\mu <m_{A'} <\,$few\,GeV, and the DM annihilation proceeds via 
creation of a pair of $A'$ particles \cite{Pospelov:2007mp}. 
Thus, we assume the stopped pion/muon source in Earth's interior to be a likely source of neutrinos. The most sensitive experiment in this energy range, $E_\nu< m_\mu/2$, is Super-Kamiokande, by virtue of its large size, low threshold and deep underground location. In this energy range, the most ``advantageous" neutrino species is $\overline{\nu}_e$, as it has the largest cross section and a more readily detectable inverse beta-decay signature.  Moreover, since the Earth's dimensions are larger than the typical neutrino oscillation length at these energies, we need to consider the production of both $\overline{\nu}_e$ and $\overline{\nu}_\mu$. 
We compute the event rate at Super-Kamiokande by using the neutrino flux from Eq.~(\ref{flux}). We consider detection of $\bar{\nu}_e$ at Super-K via the inverse beta decay $(\bar{\nu}_e + p \to n+ e)$. The production of neutrinos occurs mainly through the following channel: $\pi^+ \to \mu^+ + \nu_{\mu}$, followed by $\mu^+ \to e^++\nu_{e} + \bar{\nu}_{\mu}$, where $\bar{\nu}_{\mu}$ oscillates to $\bar{\nu}_{e}$. We use the Michel spectrum of $\bar{\nu}_{\mu}$ along with the neutrino-nucleon scattering cross-section from Ref.~\cite{Strumia:2003zx} for calculating the event-rate, which is given by
\begin{equation}
\Gamma_{\rm SK} =\int \phi_{\oplus} f_{\bar{\nu}_{\mu}} (E)\,P_{\bar{\nu}_{\mu} \to \bar{\nu}_e} \,\sigma_{\bar{\nu}_{e} p} (E)\,\epsilon(E)\,N_p\,dE\,,
\end{equation}
where $\sigma_{\bar{\nu}_{e} p} (E_e) = 9.52 \times 10^{-44}$ cm$^2$ $\left(\frac{E_e \sqrt{E^2_e-m^2_e}}{\rm MeV^2}\right)$ denotes the scattering cross-section of electron anti-neutrinos with protons~\cite{Strumia:2003zx}, and $P_{\bar{\nu}_{\mu} \to \bar{\nu}_e} \simeq 1/6$ is the oscillation probability. $\epsilon (E) \simeq 30\%$ denotes the signal efficiency and $N_p = \frac{1}{9} \left(\frac{22.5\,\rm kt}{m_p}\right)$ denotes the total number of free (hydrogen) proton targets in the fiducial volume of the Super-Kamiokande detector. We compare the event rate with the upper limit from the diffuse supernovae neutrino background 
searches, that cover the same range of energies, to derive the exclusion limits in Fig.\,\ref{lowenergy}.

In the top panels, we use the SK result with pure-water (22.5 $\times$ 2970 kton-day)~\cite{Super-Kamiokande:2021jaq}, whereas, in the bottom panels, we use the SK result with  0.01wt\% gadolinium loaded water (22.5 $\times$ 552.2 kton-day)~\cite{Super-Kamiokande:2023xup}. We probe dark matter fraction of $f_{\chi} \geq 0.1\%$ with the gadolinium loaded water result and $f_{\chi} \geq 0.01\%$ with pure-water result. The existing exclusions
from several surface and underground direct detection
searches are also shown for comparison. To adjust the experimental exclusions given for $f_{\chi} =1$ to the smaller fractions of interest here, we have applied a simplified method (by re-scaling the lower limits accordingly with keeping the ceilings fixed) as described in~\cite{McKeen:2022poo}. We note, however, this approach gives a reasonable approximation to more computationally intensive calculations in refs~\cite{Emken:2017qmp,Mahdawi:2017utm,Emken:2018run}.

\subsection{High Energy Neutrinos from 	Direct DM Annihilation}

Earth's capture of dark matter in the optically-thick regime imply the existence of a light mediator that causes the scattering cross section to be large. The annihilation of dark matter may result in the stopped meson source of neutrinos, as described in the previous subsection, but may also give rise to prompt higher-energy neutrinos. For example, high-energy neutrinos can come from a few annihilation products that decay promptly, giving continuum spectra up to
$E_{\nu} = m_{\chi}$, such as $\chi \chi \to W^+W^-, b\bar{b}, \tau \bar{\tau}$. 
It is also possible that the annihilation proceeds via an intermediate step of light mediators, {\em e.g.} 
$\chi\bar\chi \to A'A'$, with direct prompt decay of mediators to neutrinos, $A'\to \nu\bar\nu$. 
This would be the case in the model of gauged lepton number (such as  $B-L$ or $L_\mu-L_\tau$ gauge symmetries). In addition, a virtual $Z$ or $A'$ may lead to the direct annihilation of dark matter to monochromatic neutrinos, $\chi\bar\chi \to A'\to \nu\bar\nu  $.
Searches for prompt higher-energy neutrinos arising from DM annihilation in the Sun by Super-Kamiokande~\cite{Super-Kamiokande:2015xms}, ANTARES~\cite{ANTARES:2016xuh}, and  IceCube~\cite{IceCube:2016dgk,IceCube:2021xzo} yield stringent constraints on DM interactions.  Apart from these solar searches, high energy neutrinos arising from the center of the Earth has also been used to set constraints/projections on DM-nucleon scattering cross-section~\cite{ANTARES:2016bxz,Super-Kamiokande:2004pou,Mijakowski:2020qer,Hyper-Kamiokande:2018ofw,Renzi:2023pkn}.

Direct annihilation to neutrino pairs produces a line at $E_{\nu} = m_{\chi}$, which represents an interesting signature as discussed in \cite{ElAisati:2017ppn,IceCube:2021xzo,IceCube:2023ies}. In this work we aim to place a constraint on such a line signature from the Earth-bound DM annihilation by recasting the existing searches. We use the IceCube DeepCore data with a total live-time of 6.75 years for this purpose. More specifically, IceCube collaboration has searched this neutrino line in their data  with a total live-time of 6.75 years with the direction of the Sun. Given non-detection, it leads to a upper limit on dark matter annihilation rate in the mass range of $m_{\chi} = [10,100]$ GeV~\cite{IceCube:2021xzo}. We use the corresponding upper limit to derive the exclusion limits in Fig.\,\ref{highenergy}. The exclusion limits are simply derived from the fact that flux of Earth-bound DM particles can not exceed the flux upper limit: $\phi_{\oplus} \leq \Gamma^{90}_{\rm ann}/(4\pi D^2)$, where $D=1.5 \times 10^8$ km denotes the Earth-Sun distance and $\Gamma^{90}_{\rm ann}$ denotes the annihilation rate upper limit at 90\% C.L. 

We use the tabulated values of DM annihilation rate upper limit (90\% C.L.) for the DM mass range of $m_{\chi} = [10,100]$ GeV~\cite{IceCube:2021xzo}, and extrapolate it upto $m_{\chi} = 10^6$ GeV by scaling the neutrino-nucleon scattering cross-section. We did not consider $m_{\chi} \geq 10^6$ GeV as the trapping of $\chi$ particles becomes more and more inefficient with larger $m_{\chi}$ and as a consequence, it does not cover any  additional parameter space as compared to the underground detectors. We show the existing constraints from underground, surface as well as the high-altitude detectors for comparison. We found that in the regime of relatively small  $f_{\chi}$,  $f_{\chi} \leq 10^{-4}$, direct  annihilation of Earth-bound DM into neutrinos covers a part of the parameter space with 
$\sigma_{\chi n} \in  \left[10^{-26}-10^{-28}\right] \,{\rm cm}^2$, which is otherwise unexplored.

\section{ Conclusions}\label{conclusion}
A subdominant component of dark matter with large scattering cross section on nucleons represents a realistic possibility in several classes of dark sector models. In this work, we investigate generic consequences of such a scenario on the neutrino signals from annihilating dark matter.    We find that in the optically think regime, {\em i.e.} when the scattering length is much shorter than the Earth's dimensions, the accumulation inside the Earth would provide a larger neutrino flux than the Sun or other planets. If the annihilation proceeds primarily into the light mesons, one should expect a new neutrino source from stopped mesons. The neutrino signals are expected to be dominated by $\bar\nu_e$, that mostly originate from $\bar\nu_\mu\to \bar\nu_e$ oscillations. 

To limit the strength of the source, one can use existing searches of the diffuse supernova neutrino background, where record sensitivity has been achieved with the Super Kamiokande neutrino detector. The results show that neutrinos from the stopped meson source can probe abundances of strongly interacting fraction down to $f_\chi\sim 10^{-4}$. This is not as sensitive as direct annihilation to visible modes in the volume of the SK detector, as the probability of detecting a neutrino passing through the volume of a detector is quite low at these energies. At the same time, the sensitivity extends to higher range of masses, as the depletion of the surface abundance of dark matter does not affect the neutrino flux. We have also shown that if the direct annihilation to neutrinos is allowed, or a significant flux of the neutrino can be obtained from mediators decaying in flight, the sensitivity extends to higher masses of dark matter, and smaller abundances due to the rapid growth of the neutrino cross section with energy. 

 \section*{Acknowledgments}
 We sincerely thank M. Bustamante, S. Palomares-Ruiz for helpful discussions. M.P. is supported in part by U.S. Department of Energy Grant No. DE-SC0011842. M.P. is grateful to Perimeter Institute for theoretical physics for hospitality. AR acknowledges support from the National Science Foundation (Grant No. PHY-2020275), and  the Heising-Simons Foundation (Grant 2017-228).  

\bibliographystyle{JHEP}
\bibliography{ref.bib}
\end{document}